\documentclass[12pt,a4paper]{article}
\usepackage{jheppub} 

\usepackage[utf8]{inputenc}
\usepackage[T1]{fontenc}
\usepackage{lmodern}
\usepackage[british]{babel}
\usepackage{amsmath}
\usepackage{amssymb}
\usepackage{amsfonts}
\usepackage{color}
\usepackage{float}
\usepackage{stmaryrd}
\usepackage{url}
\usepackage{graphicx}
\usepackage{epstopdf}
\usepackage{placeins,bbm}
\newcommand{\beq}{\begin{equation}}
\newcommand{\eeq}{\end{equation}}
\newcommand{\bea}{\begin{eqnarray}}
\newcommand{\eea}{\end{eqnarray}}
\newcommand{\nn}{\nonumber}
\newcommand{\re}{\mathrm{Re}}

\newcommand{\de}{\mathrm{d}}
\def\bx{{\mathbf{x}}}

\def\lsi{\raise0.3ex\hbox{$<$\kern-0.75em\raise-1.1ex\hbox{$\sim$}}}
\def\gsi{\raise0.3ex\hbox{$>$\kern-0.75em\raise-1.1ex\hbox{$\sim$}}}

\newcommand{\tr}{\ensuremath{\mathrm{tr}}}

\providecommand{\href}[2]{#2}

\newcommand{\Teffc}{(T_c)_{\text{eff}}}

\newcommand{\erw}[1]{\langle #1 \rangle}
\newcommand{\eqa}[1]{eq.~(\ref{#1})}
\begin{document}
\title{Effective lattice Polyakov loop theory vs.~full $SU(3)$ Yang-Mills at finite temperature}
\author{G.~Bergner$^1$, J.~Langelage$^2$,  O.~Philipsen$^1$}
\affiliation{$^1$ Institut f\"ur Theoretische Physik, Goethe-Universit\"at Frankfurt,\\
Max-von-Laue-Str.~1, 60438 Frankfurt am Main, Germany}
\affiliation{$^2$ Institute for Theoretical Physics, ETH Z\"urich, CH-8093 Z\"urich, Switzerland}

\date{}
\abstract{
A three-dimensional effective theory of Polyakov loops has recently been derived from Wilson's Yang-Mills
lattice action by means of a strong coupling expansion. It is valid in the confined phase up to the 
deconfinement phase transition, for which it predicts the correct order and gives quantitative estimates
for the critical coupling.
In this work we study its predictive power for further observables like
correlation functions and the equation of state.
We find that the effective theory correctly reproduces qualitative features
and symmetries of the full theory as the continuum is approached. Regarding quantitative predictions,
we identify two classes of observables by numerical comparison as
well as analytic calculations: correlation functions and their associated mass scales
cannot be described accurately from a truncated effective theory, due to its inherently non-local nature
involving long-range couplings.
On the other hand, phase transitions and bulk thermodynamic quantities are accurately reproduced
by the leading local part of the effective theory. In particular, the effective theory description is numerically superior when computing the equation of state at low temperatures or the properties of the phase 
transition.
}
\maketitle
\section{Introduction}

The use of Polyakov loop models as a simplified effective description of the pure glue sector of
QCD at finite temperature has a long history. 
This is based on the expectation that, around the deconfinement phase transition, the dynamics of Yang-Mills theory is governed by the degrees of 
freedom which also constitute the order parameter for the global symmetry breaking driving the transition.
Once an appropriate model is at hand, it is easier to analyse than the original theory, both with analytic 
methods or with numerical simulations. The goal is to obtain an effective description of 
Yang-Mills theory and ultimately full QCD, which would allow to determine the 
phase diagram and physical properties
of QCD at finite baryon density, where lattice QCD exhibits its sign problem. For a recent example 
and references see \cite{models}.
In more recent approaches the aim is to actually derive 
the effective Polyakov loop theory directly from Yang-Mills or QCD by  
perturbation theory \cite{vuorinen}, strong coupling expansions \cite{Langelage:2010yr} (see also \cite{Christensen:2013xea}, where the same method has been applied to large $N_c$),
Monte Carlo methods \cite{jena,green1,green2} or the functional renormalisation group \cite{janp}. 

The various techniques employed to construct the effective theory each have their advantages
and disadvantages.
Effective theories derived by weak or strong coupling methods are only valid in the deconfined or confined phase, respectively, and thus are complementary. Their advantage are analytic
expressions between the effective and fundamental couplings, which make for economic and 
flexible use of the effective theory to arrive at predictions. The disadvantage is the systematic
error introduced by truncating the expansions at finite order. 
Non-perturbative approaches, on the other hand, have the advantage to potentially work at all couplings
and to give a valid description on both sides of the phase transition. However, in this case the couplings
are only known numerically and have to be recomputed for every change in the parameters of the
original theory. Moreover, any particular form of an effective 
action with a finite number of terms necessarily implies truncations in the space of effective couplings, 
and an estimate of the implied systematic error is often more difficult than in series expansions.

This paper is devoted to a study of the systematics of a three-dimensional effective lattice action for Yang-Mills theory derived from the four-dimensional Wilson action by the strong coupling expansion 
\cite{Langelage:2010yr}. The one-coupling effective theory derived in that work gives the correct prediction
for the order of the $SU(2),SU(3)$ deconfinement transitions as well as the corresponding critical
temperature $T_c$ to about 10\% accuracy in the continuum limit.
Here we extend the comparison between the effective and full theory to correlation functions of
Polyakov loops, i.e.~the free energy of a static quark anti-quark pair, 
as well as thermodynamic functions. 

This work is organized as follows. In the next section we summarise the derivation of the effective theory.
It correctly reproduces all qualitative features and symmetries of the full theory as the continuum is approached.
We consider two classes of observables: In Section \ref{sec:PLLOOP} correlation functions and the associated length or mass scales.
We find that they cannot be predicted accurately from a truncated effective theory, due to the inherent sensitivity to long-range interactions.
On the other hand, phase transitions and bulk thermodynamic quantities in Section \ref{sec:THPOT} are accurately reproduced
by the leading local part of the effective theory. 

\section{The effective lattice Polyakov loop theory}

The effective lattice Polyakov loop theory is defined starting from Wilson's lattice Yang-Mills action 
on a $N_s^3\times N_\tau$ lattice by splitting the link integrations into a spatial and temporal part,
\bea
	Z&=&\int [\de U_\mu] e^{-S_{YM}[U]}
		\equiv\int [\de W]\,e^{-S_\text{eff}[W]}\;,\nn\\
	S_{\mathrm{eff}}[W]&=&-\sum_{i=1}^\infty\bar{\lambda}_i(\beta,N_\tau)S_i[W]  \;.	
\eea
The individual terms in the effective action, $S_i[W]$, depend on temporal Wilson 
lines, $W(\bx)=\prod_{\tau=1}^{N_\tau}U_0(\bx,\tau)$, or their traces, the Polyakov loops $L_i=\tr[W(\bx_i)]$, which are the remaining integration variables in the path integral.
Note that, without truncations, the effective action is unique and exact. 
Since all spatial links, which 
are originally coupled by nearest neighbour interactions, were
integrated over, the effective action is of long-range type, irrespective of the way it is determined.
It contains interactions of Polyakov lines at all distances, even a non-local form is allowed.
Here we consider the case where the Boltzmann factor is expanded 
in a strong coupling expansion so that all link integrations can be performed analytically. 
The leading terms in the infinite volume limit read
\bea	
	S_{\mathrm{eff}}[W]
	&=&-\sum_{<i,j>}\ln(1+2\lambda_1(\beta,N_\tau) {\rm Re} L_i L^\dag_j )
	-\sum_{[k,l]}\ln(1+2 \lambda_2(\beta,N_\tau) 
	{\rm Re} L_k L^\dag_l ) \nn\\
        &&-\sum_{<<k,l>>}\ln(1+2 \lambda_3(\beta,N_\tau) 
	{\rm Re} L_k L^\dag_l )+\ldots\;,
\label{eq:efft}	
\eea
where $<ij>$ denotes all pairs of nearest neighbours in the first term, $[kl]$ all pairs of
next-to-nearest neighbours with distance $R/a=\sqrt{2}$, and  $<<kl>>$ all pairs with distance $R/a=2$.
Without truncations, the action consists of infinitely many, generically not bilinear, terms
with Wilson lines to all powers, all distances and in all representations, where the latter are a remnant of our preferred computational method, the character expansion. The higher representations can be converted back into products of the fundamental one, i.e.~we may choose to work solely with fundamental loops to arbitrary powers.
Note that in eq.~(\ref{eq:efft})
we have resummed higher powers of nearest neighbour interactions and next-to-nearest neighbours  
into a logarithm. This summation of an infinite number of terms redefines the couplings and improves the convergence behaviour, as discussed in detail in section \ref{eos}.

Using the strong coupling expansion, the terms in the effective action
are naturally ordered by the lowest power of $\beta$ at which the corresponding effective coupling enters. Usually we express the effective couplings in terms of the fundamental character expansion coefficient $u=u(\beta)=\beta/18+O(\beta^2)$, which shows better convergence. 
The relation between $u$ and $\beta$ can be computed to arbitrary precision, 
hence we can use them synonymously.
A complete list of the couplings used in this work is summarized in the appendix, \eqa{eq:couplings1} and \eqa{eq:couplings2}.
As these expressions show, higher order couplings are parametrically suppressed with growing $N_\tau$, which corresponds to finer lattices at fixed temperature $T=1/(aN_\tau)$.

Regardless of the truncation, the effective theory exhibits centre symmetry by construction and its spontaneous breaking
at finite temperature. 
In \cite{Langelage:2010yr} it was found that the theory truncated to the leading nearest neighbour interaction correctly predicts the different orders of the deconfinement phase 
transition for $SU(2)$ and $SU(3)$. Moreover, its predicted critical couplings for the phase transition
agree with those from Monte Carlo simulations of the full theory to better than 10\% accuracy
for a wide range of temporal lattice sizes, $N_\tau\leq 16$, cf.~table \ref{tab:su3_betas}. With an appropriate scale setting by means
of a known 4d beta-function \cite{Necco:2001xg}, this permits the calculation of the deconfinement temperature $T_c$
in the continuum limit with similar accuracy \cite{fromm}. In the following sections we investigate the
predictive power of the effective theory for correlation functions and bulk thermodynamic quantities.
\begin{table}[t]
\begin{center}
\begin{tabular}{|c||c|c|c|c|}
\hline
	$N_\tau$ &$\lambda_1$  & $(\lambda_1,\lambda_2)$ &
	 $\mbox{4d YM}$  \\
\hline
2   &  5.1839(2)   & 5.0174(4)      &  5.10(5)     \\
3   &  5.8488(1)   & 5.7333(3)      &  5.55(1)     \\
4   &  6.09871(7)  & 6.0523(1)     &  5.6925(2)   \\
6   &  6.32625(4)  &  6.32399(3)   &  5.8941(5)  \\
8   &  6.43045(3)  &  6.43033(2)  &  6.001(25)  \\
10  &  6.49010(2)  &  6.49008(2)  &  6.160(7)   \\
12  &  6.52875(2)  & 6.52874(1)   &  6.268(12)  \\
14  &  6.55584(2)  & 6.55583(1)   &  6.4488(59)  \\
16  &  6.57588(1)  & 6.57587(1)   &  6.5509(39) \\
\hline
\end{tabular}
\caption{Critical couplings $\beta_c$  for $SU(3)$ for the one- and two-coupling effective 
theories compared to simulations of the 4d theory \cite{fingberg_heller_karsch_1993, Kogut_et_al_1983,Francis:2013cva}.
}
\label{tab:su3_betas}
\end{center}
\end{table}
%

\section{Polyakov loop correlators and static quark free energy}
\label{sec:PLLOOP}
\subsection{Two-point correlators}

The spectrum of a theory is encoded in its correlation functions. A natural testing ground for the effective action are thus Polyakov loop correlators. Their exponential decay
represents the (unrenormalised) free energy of a static quark anti-quark pair \cite{svet}, 
\begin{equation}
\langle L(R)L^\dag(0)\rangle=e^{- F(R,T)/T}\;.
\end{equation}
 \begin{figure}
\includegraphics[width=0.5\textwidth]{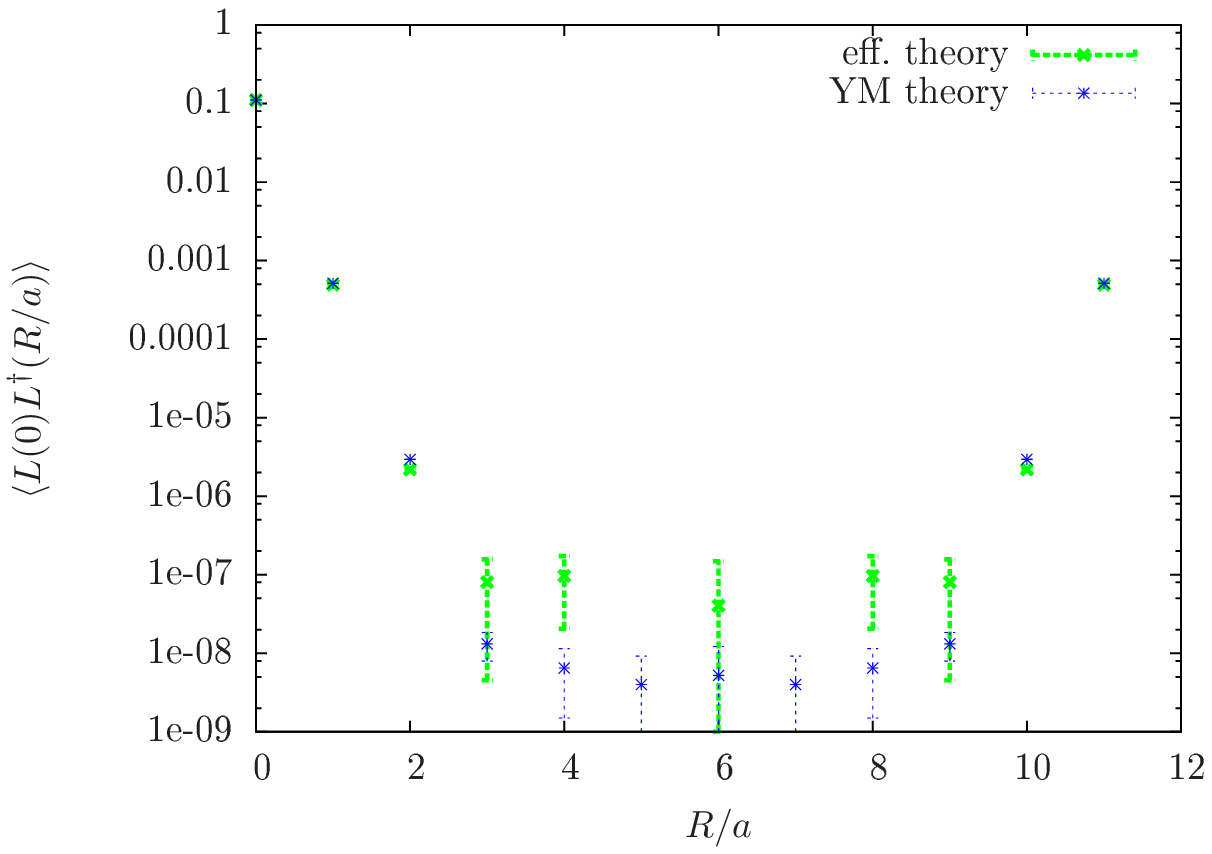}
\includegraphics[width=0.5\textwidth]{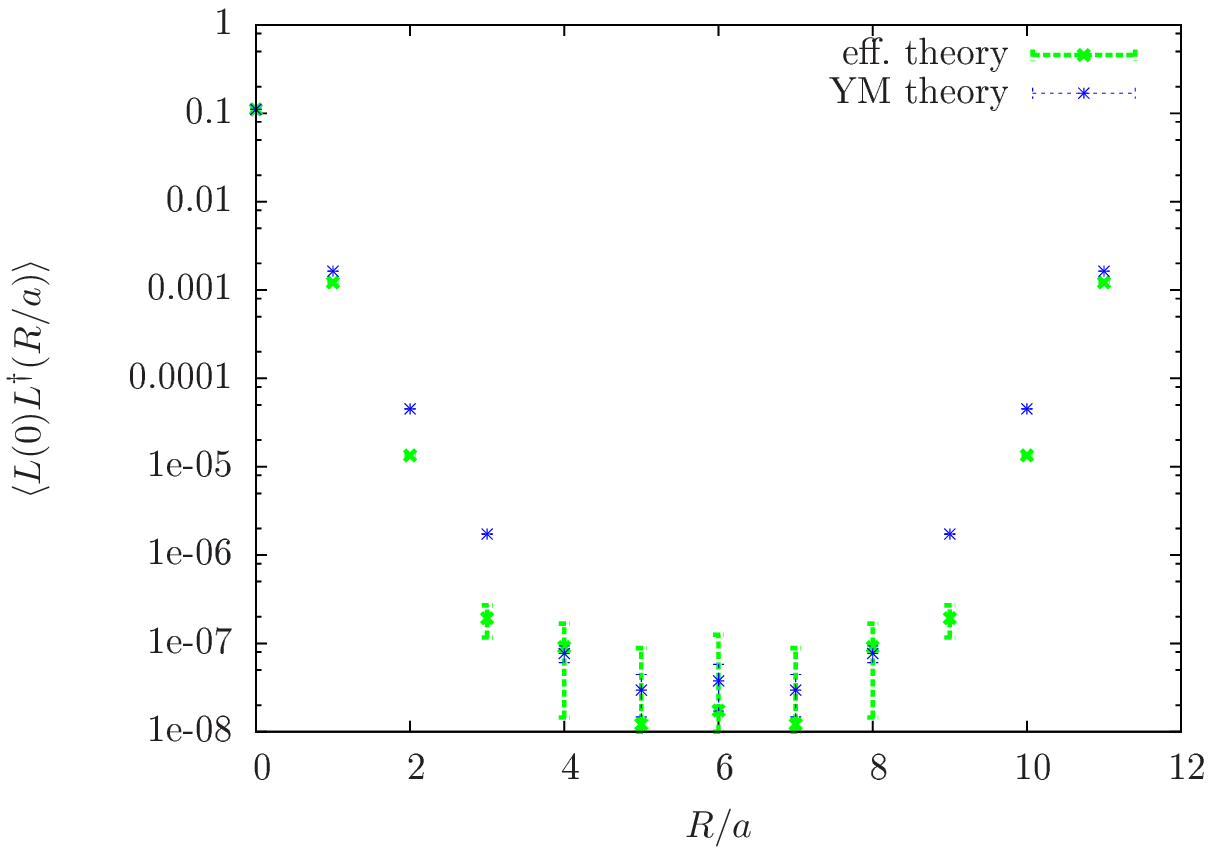}
 \caption{Comparison of the Polyakov loop correlator between the one-coupling effective theory and full Yang-Mills theory at $\beta=5.0$ (left) and $\beta=5.4$ (right). 
 Both simulations were done on a $6\times 12^3$ lattice.
}
 \label{fig:basicc}
\end{figure}
A direct comparison of on-axis correlators between the one-coupling effective theory and full Yang-Mills is
shown in figure \ref{fig:basicc} for two different values of the lattice coupling $\beta$. Note that in the full
Yang-Mills simulation an algorithm for exponential error reduction \cite{lw} was employed, whereas the 
data for the effective theory have been obtained only with the comparably small improvement of a multi-hit algorithm. Quantitative agreement
is observed for short lattice distances $R/a=0-2$ for $\beta=5.0$, while the effective theory data still follow the general shape of the full correlator, but start to quantitatively deviate for 
$R/a\geq2$ at the larger coupling $\beta=5.4$. 

\begin{figure}[t]
 \includegraphics[width=0.5\textwidth]{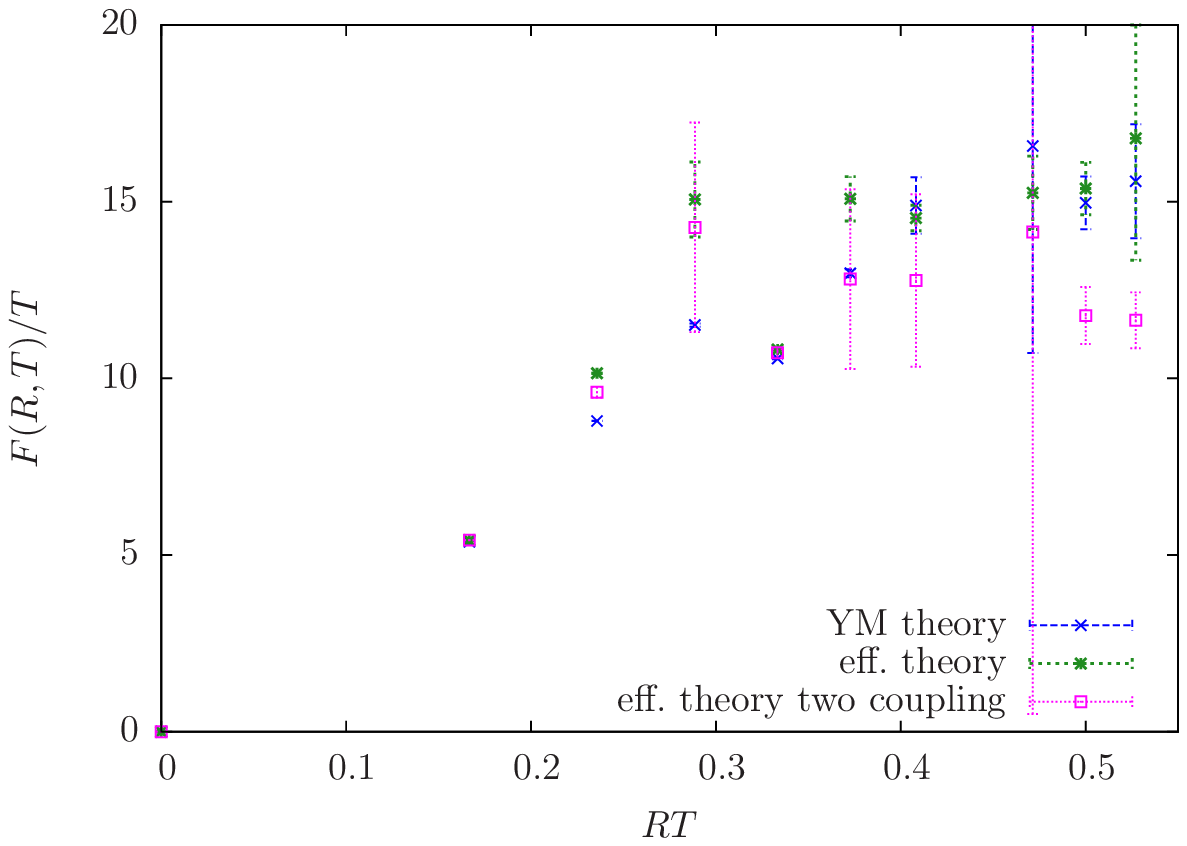}
\includegraphics[width=0.5\textwidth]{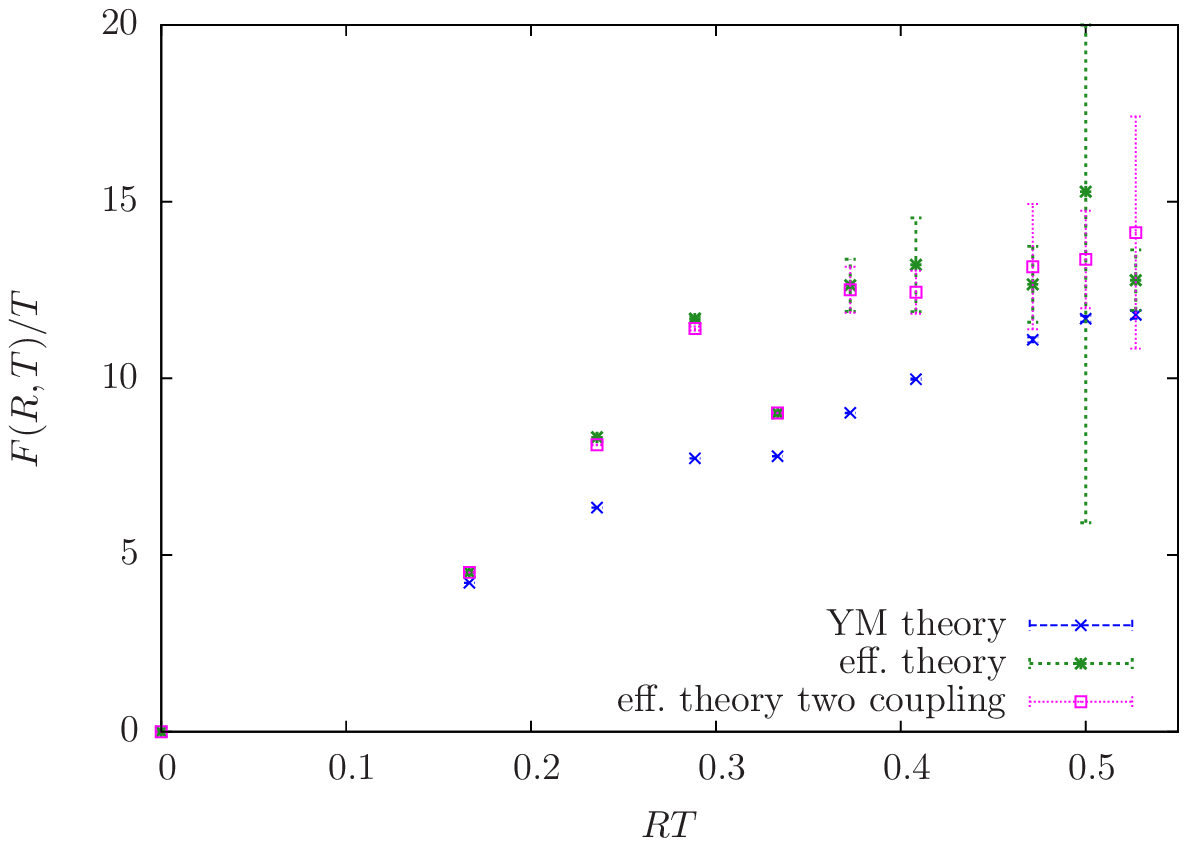}
\vspace*{-0.7cm}
 \caption[]{Free energy of a static quark-antiquark pair for full Yang-Mills and the effective theory 
 with one and two coupling constants at $\beta=5.0$ (left) and $\beta=5.4$ (right) on $12^3\times 6$. 
 This comparison includes also off-axis correlators.}
 \label{fig:FRT}
\end{figure}

\begin{figure}[t]
\includegraphics[width=0.5\textwidth]{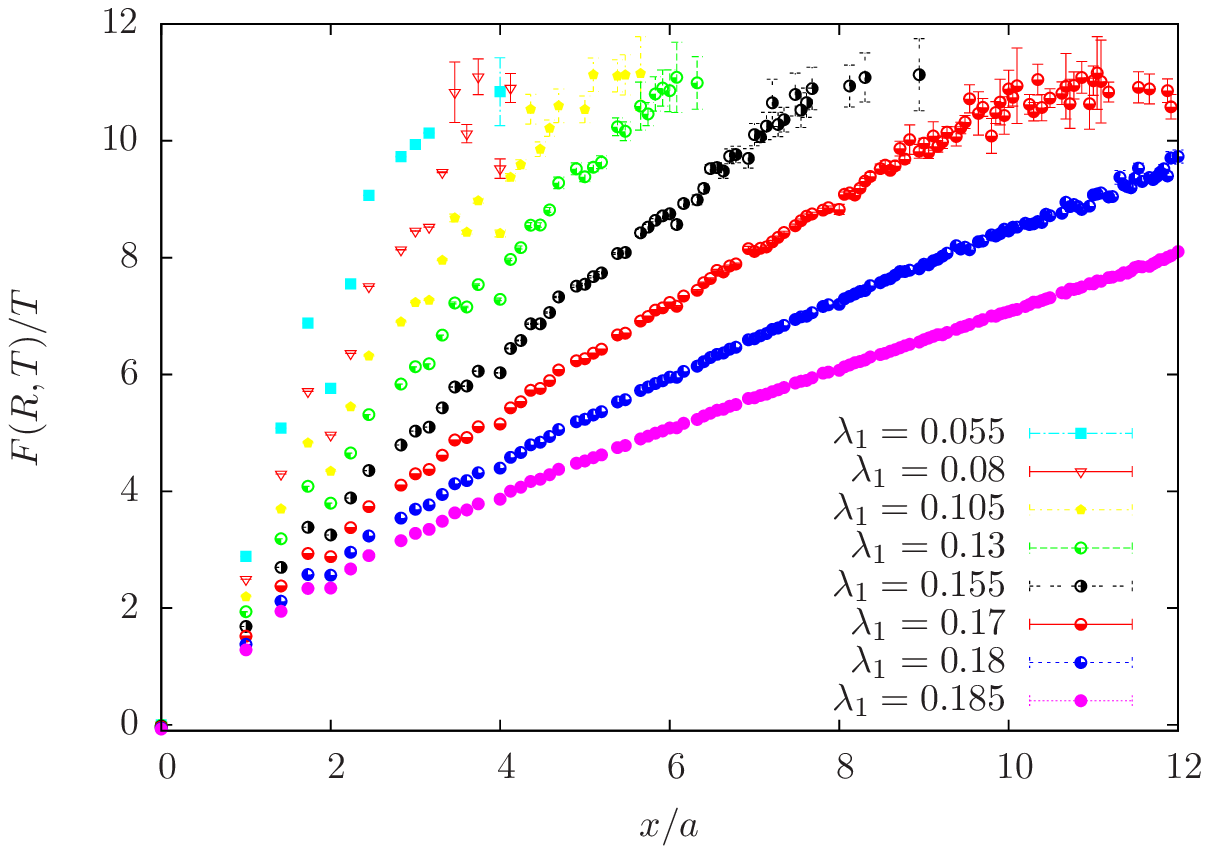}
\includegraphics[width=0.5\textwidth]{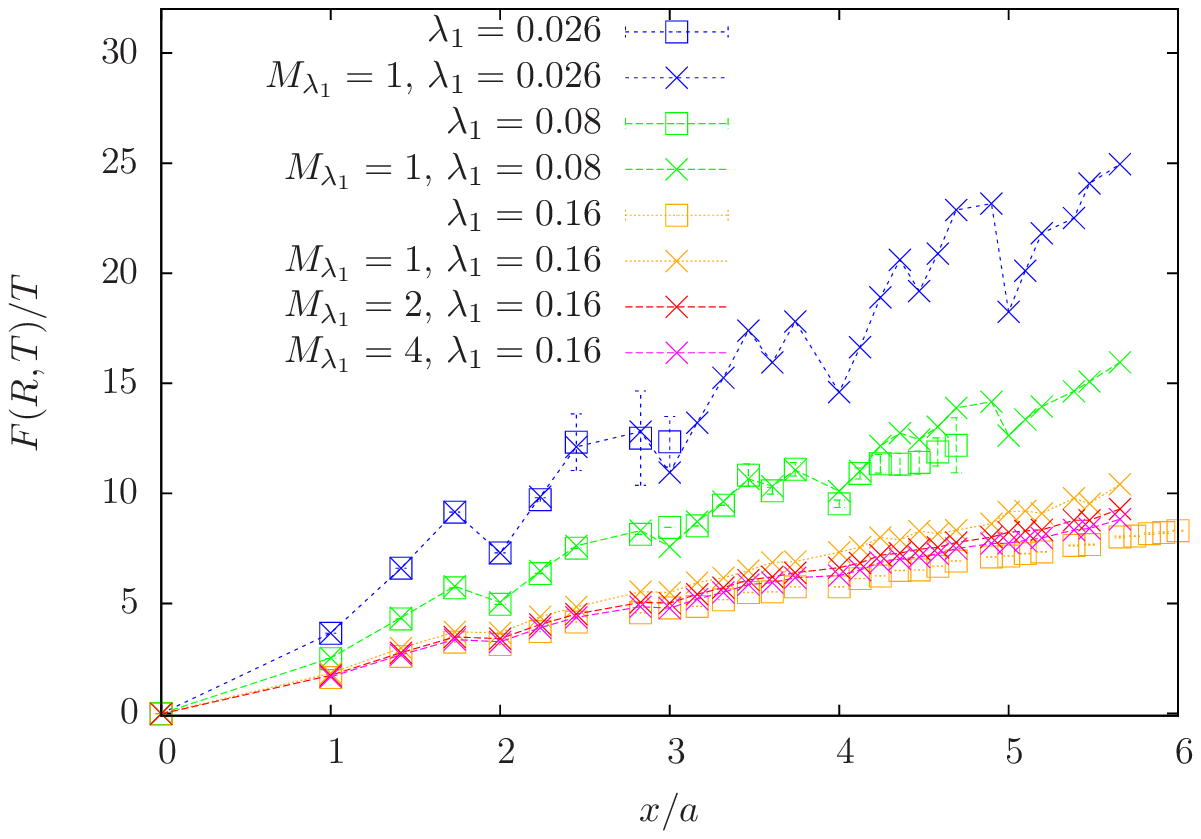}
\vspace*{-0.5cm}
 \caption[]{Left:  Free energy for different values of $\lambda_1$ in the effective theory on a $32^3$ lattice. 
 The rotational invariance is restored at larger values of the coupling.
Right: Numerical simulations
are compared to analytic results of the small $\lambda_1$ expansion, eq.~(\ref{eq:smallapprox}). 
}
 \label{fig:effrest}
\end{figure}

Note that the correlators are systematically smaller, i.e.~the corresponding free energies
are larger in the effective theory. This is shown in continuum units in figure \ref{fig:FRT},
with deviations increasing with distance. The off-axis correlators are also included in this figure.
These are a measure for the breaking of rotational invariance by the lattice discretisation, which appears to be amplified in the effective theory compared to the full Yang-Mills theory. This effect gets alleviated when also the next-to-nearest neighbour coupling is included in the
effective theory. However, the improvement is small because of the smallness of the effective coupling,
with values of $\lambda_2(\beta=5.0)=1.9\times 10^{-5}$ and $\lambda_2(\beta=5.4)=5.6\times 10^{-5}$, when using its strong coupling expansion eq.~(\ref{eq:couplings2}). On the other hand, when the effective coupling $\lambda_1$ (or equivalently $\beta$) is
raised, rotational invariance gets restored also in the one-coupling theory, as figure \ref{fig:effrest} (left)
illustrates, i.e.~the effective action eventually reproduces the continuum symmetries. 
In this regime of larger $\lambda_1$ just below its critical value, which corresponds to larger $\beta$ and hence finer lattices, it is also possible to distinguish between a linear part at large distances and a Coulomb part at short distances. 

\subsection{Weak coupling expansion in the effective theory}

Because of the smallness of the effective couplings, it is natural to consider perturbation theory for the effective action. Indeed, the correlator of Polyakov loops can be expressed as a power series in the coupling truncated at $M_{\lambda_1}$, 
\beq
\langle L(R)L^\dag(0)\rangle\approx \sum_{n=1}^{M_{\lambda_1}} N_n(R/a)\lambda_1^{l_n(R/a)}\;.
\label{eq:smallapprox}
\eeq 
If the distance $R/a$ is greater than zero,  $l_1$ is
the ``taxi-driver'' distance on the lattice, i.e.~the minimal number of links connecting
the two correlated points. The coefficient functions $N_n$ count the number of possible paths of length $l_n$ that connect the correlated points. They have been estimated with a numerical algorithm 
and are summarized in Table \ref{tab:lapprox}.
\begin{table}[t]
\begin{center}
 \begin{tabular}{|c|c|c|c|c|c|c|c|c|}
 \hline
 $R/a$&$N_1$&$l_1$&$N_2$&$l_2$&$N_3$&$l_3$&$N_4$&$l_4$\\
 \hline
 0 & 1 & 0 & 0 &  0 &  0 & 0 & 24 & 4 \\
1 & 1 & 1 & 4 & 3 & 8 & 4 & 76 & 5\\
1.41421 & 2 & 2 & 18 & 4 & 12 & 5 & 316 & 6\\
1.73205 & 6 & 3 & 60 & 5 & 54 & 6 & 1128 & 7\\
2 & 1 & 2 & 12 & 4 & 8 & 5 & 240 & 6\\
2.23607 & 3 & 3 & 49 & 5 & 22 & 6 & 909 & 7\\
2.44949 & 12 & 4 & 178 & 6 & 98 & 7 & 3648 & 8\\
2.82843 & 6 & 4 & 148 & 6 & 44 & 7 & 2918 & 8\\
3 & 1 & 3 & 24 & 5 & 8 & 6 & 588 & 7\\
3.16228 & 4 & 4 & 108 & 6 & 30 & 7 & 2398 & 8\\
3.31662 & 20 & 5 & 444 & 7 & 158 & 8 & 10160 & 9\\
3.4641 & 90 & 6 & 1872 & 8 & 720 & 9 & 43236 & 10\\
3.60555 & 10 & 5 & 361 & 7 & 74 & 8 & 8253 & 9\\
3.74166 & 60 & 6 & 1524 & 8 & 472 & 9 & 36242 & 10\\
4 & 1 & 4 & 40 & 6 & 8 & 7 & 1260 & 8\\
\hline
\end{tabular}
\caption{A table of the expansion coefficients of the correlation function \eqa{eq:smallapprox}. Only the shortest distances are listed in this table.}
\end{center}
\label{tab:lapprox}
\end{table}
Figure \ref{fig:effrest} (left) illustrates how for small values of 
$\lambda_1$, i.~e.\ at strong coupling, this analytic result fully reproduces the numerical simulations of the effective model. With growing values of $\lambda_1$ and larger distances, higher orders in the expansion become important. 
At the largest $\lambda_1$ value shown in figure \ref{fig:effrest}, which corresponds at $N_\tau=4$ to $\beta\approx 6$, the fourth order expansion is still a good approximation for the short distance correlation.

\subsection{Effective $T$-dependent string tension}

\begin{figure}
\includegraphics[width=0.5\textwidth]{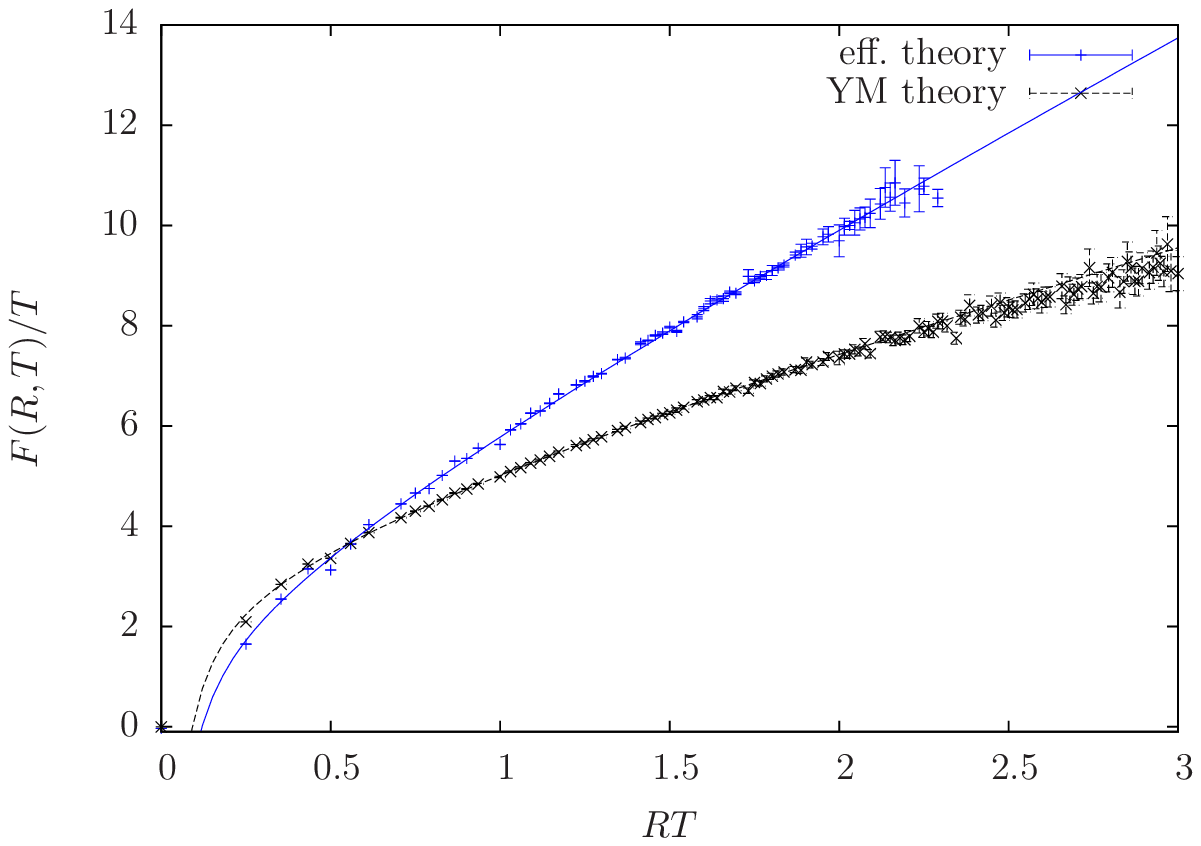}
\includegraphics[width=0.5\textwidth]{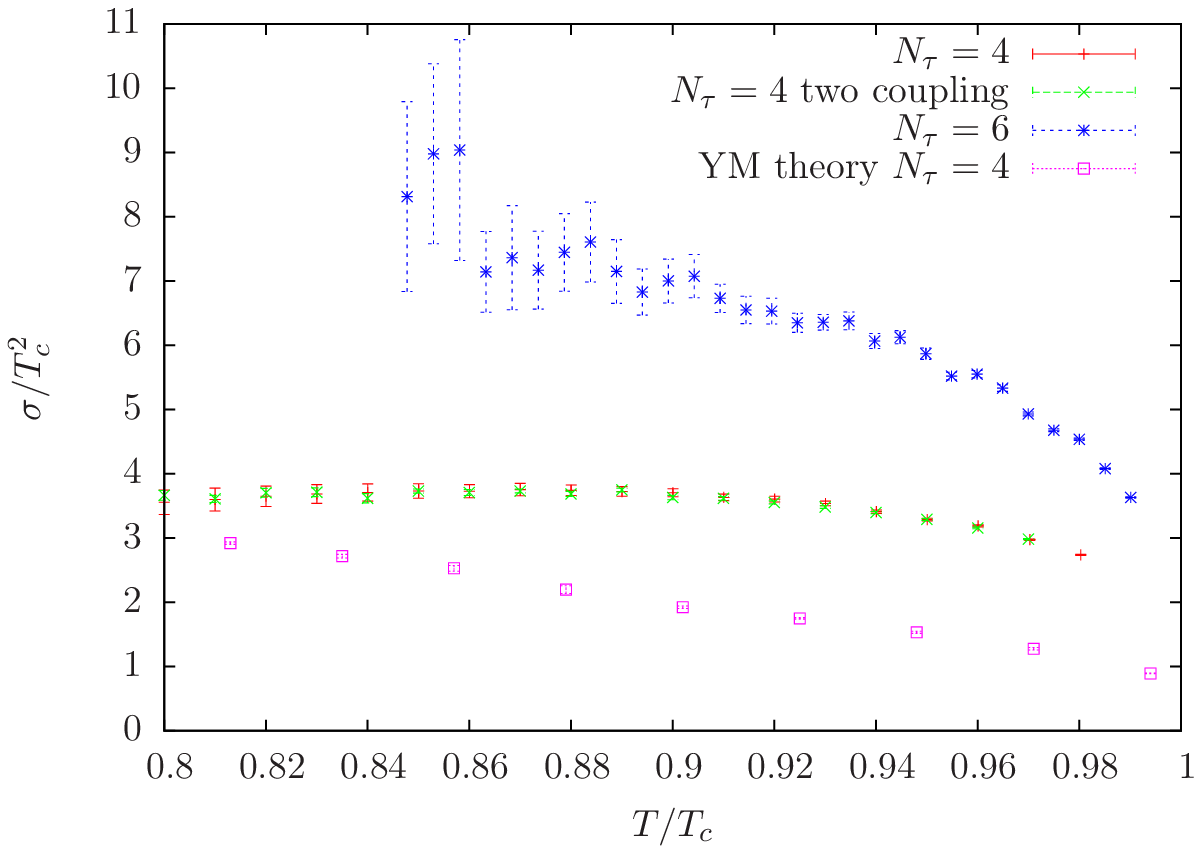}
\vspace*{-0.7cm}
 \caption[]{Left:
Free energy from the effective one-coupling theory 
at $T=90\%\Teffc$ on $32^3\times 4$ compared to Yang-Mills theory. 
The error bars include a systematic error as the difference between the $O(u^{10})$ and $O(u^9)$ truncation of  eq.~(\ref{eq:couplings1}). Right: Effective temperature dependent string tension. The 
Yang-Mills data is taken from \cite{Cardoso:2011hh}.
}
 \label{fig:string}
\end{figure}

The linear part of the free energy corresponds to a temperature dependent effective string tension, 
which arises from the Boltzmann average over the linear pieces of the excitations of the static potential.
This string tension decreases with $\lambda_1$ (or $\beta$), which is tantamount to increasing 
temperature at fixed $N_\tau$, in accord with full Yang-Mills theory 
\cite{Kaczmarek:1999mm,Cardoso:2011hh}. To make the comparison quantitative, we fit our correlator in continuum units with the same
ansatz used in \cite{Kaczmarek:1999mm}  (details of the functional form are inspired by string models valid at large distances), 
\bea
\frac{F(R,T)}{T}&=&v_0+\frac{1}{2}\ln(1+(2RT)^2)+
\left[\frac{\pi}{12}-\frac{1}{6}\arctan(2RT)\right] \frac{1}{RT}\nn\\
&&+\left[\frac{\sigma(T)}{T^2}-\frac{\pi}{3}+\frac{2}{3}\arctan\left(\frac{1}{2RT}\right)\right]RT \;.
\eea
At short distances $R<T^{-1}$ the temperature effects disappear and the free energy is dominated by
its ground state, the static potential of the vacuum,
\beq
 F(R,T)\stackrel{R\rightarrow 0}{\longrightarrow}V(R)=c_1+\frac{c_2}{R}+\sigma R\;.
 \eeq
Data from the effective theory close to the phase transition, $T=0.90 T_c$ are shown in figure \ref{fig:string} (left). They are well described by the finite temperature fit in the long distance region while being also compatible with the corresponding vacuum fit function appropriate for short distances.
Thus, the qualitative features of the free energy of Yang-Mills theory are reproduced. However, the numerical values for the temperature dependent string tension for $N_\tau=6$ are significantly overestimated.
A collection of fit results in comparison with the full answer is shown in figure \ref{fig:string} (right).
We see that on $N_\tau=4$ the results are closer to the true answer and that the effective theory prediction
seems to grow with $N_\tau$. 

This incorrect scaling behaviour of the effective string tension is an artefact of the one-coupling theory. For all parirings $(\beta,N_\tau)$ with constant effective coupling $\lambda_1(\beta,N_\tau)$ 
the correlators as a function of distance in lattice units are the same. At large $R$ this implies that for two different $N_\tau$ with the same value of $\lambda_1$ the string tensions are related by
\begin{equation}
 \left(\frac{\sigma(T_1)}{T_1^2}+\frac{\pi}{3}\right)\frac{1}{(N_\tau)_1}=\left(\frac{\sigma(T_2)}{T_2^2}+\frac{\pi}{3}\right)\frac{1}{(N_\tau)_2}\;,
\end{equation}
where the temperatures are determined by the corresponding $N_\tau$ and $\beta$.
(The scale is set
such that $T((\lambda_1)_c)=T_c$ for all $N_\tau$).
This forces the string tension 
to scale approximately with $N_\tau$ in the region close to $T_c$.
The solution is an effective theory with more than one coupling constant. Then there is a critical 
(hyper-)surface and at each $N_\tau$ the phase transition can occur at a different values of the coupling constants. Our strong coupling result for the next-to-nearest neighbour interaction alone is however too small for a significant change of the string tension close to the phase transition.

One can understand qualitatively, why 
the value of the string tension cannot be correctly predicted by an effective theory with only a few couplings.
We have already seen that the correlators are increasingly underestimated as the correlation
distance in lattice units grows. A fixed distance in physical units contains more and more lattice spacings
as the lattice gets finer. Correspondingly, the effective theory with a fixed number of couplings 
covers an ever smaller contribution to the correlator at fixed distance correctly. While the higher order
couplings do become parametrically smaller $\sim u^{nN_\tau+m}$, eq.~(\ref{eq:couplings1}),  
a rapidly increasing number
of them
contributes to a correlator at distance $R/a$.
Moreover, their apparent suppression with $N_\tau$ cancels in the contribution to the free energy. 
This can be demonstrated rather precisely by considering the strong coupling expansion of the 
vacuum string tension, i.e.~the limit $N_\tau \rightarrow \infty$ at fixed $\beta$. 
The expansion starting from
the 4d Yang-Mills theory is well known \cite{Montvay:1994cy} (for a finite temperature version, see \cite{Green:1982hu}),
\beq
a^2\sigma|_\textrm{YM}=-\ln u - 4u^4-12u^5+10u^6+O(u^7)
\eeq
On the other hand, expanding the 3d effective theory in powers of the
effective coupling constant $\lambda_1$ we obtain from the on-axis correlator 
\begin{equation}
 a^2 \sigma|_\textrm{eff.th.}=-\frac{1}{N_\tau}\ln (\lambda_1)-\frac{2\lambda_1}{N_\tau}+\ldots
 =-\ln u-4u^4-12 u^5+14u^6+O(u^8)\; .
\end{equation}
Only the first two terms agree with the expansion of the full theory. Note how the $N_\tau$-dependence of $\lambda_1$ is cancelled by an explicit $N_\tau$-dependence, while the second term is, as 
$\lambda_1$ itself, exponentially suppressed for increasing $N_\tau$. If we also include $\lambda_3$, the coupling for on-axis neighbours at 
distance $R/a=2$, we
obtain instead
\bea
  a^2 \sigma|_\textrm{eff.th.}&=&-\frac{1}{N_\tau}\ln 
(\lambda_1)-\frac{2\lambda_1^2}{N_\tau}-\frac{\lambda_3\lambda_1^{-2}}{N_\tau}
+\sum_{n=2}^{\frac{1}{2}(R/a-1)}c_n \lambda_3^n \lambda_1^{-2n}\ldots\nn\\
&=&-\ln u-4u^4-12 u^5+10u^6+O(u^8)\;.
\eea
The detailed form of the last term in the first line depends 
on whether $R/a$ is even or odd, but the number of terms in the sum 
scales with $R/a$. Because of the $N_\tau$-dependence of the couplings, the second term is less 
significant than the third one for larger $N_\tau$.
The leading contribution of the third term is $4u^6$, such that the
string tension is now correctly reproduced through order $u^6$. Correspondingly, the coefficients of
higher orders receive more and more contributions from long-range couplings.

We conclude that the long-range interactions may not be neglected in the 
computation of correlation functions within the effective theory. Contrary to the effective couplings themselves, 
their contribution to correlators are {\it not} suppressed by $N_\tau$ and without them the 
coefficients of the strong coupling expansion of correlators are incomplete.
While we have used the strong coupling expansion to show this,
we stress that the conclusion is independent of the way the effective theory is determined or used,
and in complete agreement with the non-perturbative observations made in \cite{green2}.

\section{Thermodynamic potentials and phase transitions \label{eos}}
\label{sec:THPOT}
\subsection{The equation of state}

In this section we test the description of bulk thermodynamic quantities by the effective theory, which are all derived from 
the partition function directly.
The fundamental ingredient to the equation of state is the free energy density in units of the temperature. In homogeneous systems it is related to the pressure $p$ as
\begin{equation}
  \frac{f}{T^4}=\frac{-p}{T^4}=-\frac{\ln Z}{VT^3}\;.
\end{equation}
For the correct renormalisation the divergent zero temperature part has to be subtracted.

In order to judge the quality of the effective theory, it is again instructive to consider the strong coupling expansion for different versions of the action,
\begin{eqnarray}
\mbox{4d YM:}&&\phantom{f(\lambda_1(u,N_\tau))} f(u,N_\tau)=-\frac{6}{N_\tau}u^{4N_\tau}+\ldots\;,\\
\mbox{eff. theory, linear action:}&&\phantom{ f(u,N_\tau)} f(\lambda_1(u,N_\tau))=-\frac{3}{N_\tau}\lambda_1^2+\ldots=-\frac{3}{N_\tau}u^{2N_\tau}+\ldots\;,\nonumber\\
\mbox{eff.theory, log. action:}&&\phantom{ f(u,N_\tau)} f(\lambda_1(u,N_\tau))=-\frac{6}{N_\tau}\lambda_1^4+\ldots=-\frac{6}{N_\tau}u^{4N_\tau}+\ldots\;.\nn
\end{eqnarray}
Here the full action is the ordinary strong coupling expansion without the detour of the effective theory \cite{Langelage:2008dj, Langelage:2010yn}, the log.~action corresponds to the first term of 
eq.~(\ref{eq:efft}) and the linear action to its leading term in $\lambda_1$ only. The explicit comparison reveals that
the resummation of higher power terms into the logarithm is necessary to correctly reproduce the leading
term of the full theory.

In a lattice simulation all expectation values are normalised on the partition function, which thus cannot
be calculated directly. The free energy density is computed indirectly
through its derivative with respect to the coupling constant $\beta$, which then has to be integrated over
\cite{Engels:1981qx,engels2,Boyd:1996bx},
\begin{equation}
 \left. \frac{f}{T^4}\right|_{\beta_0}^{\beta}=-\int_{\beta_0}^{\beta} d\beta' \Delta S(\beta')\; ,
\end{equation}
with the interaction measure 
\begin{equation}
 \Delta S(\beta)= \frac{1}{T^4}\left(\left. \frac{T}{V}\frac{d\ln Z}{d\beta}\right|_{T}-\left. \frac{T}{V}\frac{d\ln Z}{d\beta}\right|_{T=0} \right)
             =\frac{6 N_\tau^4}{N_c}\left(\erw{\re P}|_{T}-\erw{\re P}|_{T=0}\right)\; .
\end{equation}
Thus, all information of the equation of state is encoded in $\Delta S$. Computationally, this is a simple subtraction of two plaquette ($P$) expectation values averaged over all orientations and volume. Finite temperature $T$ and $T=0$ are represented in terms of a $N_\tau\times N_s^3$ and a $N_s^4$ lattice with $N_s=4 N_\tau$.

The strong coupling expansion for $\Delta S$ can be obtained from the series for the pressure \cite{Langelage:2010yr},
\begin{equation}
 \Delta S=N_\tau^4\frac{d(a^4p)}{d \beta}=N_\tau^4  K(u,N_\tau) \frac{du}{d\beta}\; ,
 \label{eq:ssc}
\end{equation}
where for $N_\tau=2,4$ we have
\begin{align}
 K(u,N_\tau=4)&=
24 u^{15}
+1458 u^{17}
-5643 u^{18}
+9945 u^{19}
-\frac{201285 }{4}u^{20}\nonumber\\
&+\frac{360638553}{1024}u^{21}
-\frac{8627830587 }{10240}u^{22}
+\frac{6648458901 }{5120}u^{23}\; ,\nn\\
K(u,N_\tau=2)&=24 u^7+270 u^9-1485
   u^{10}+3315 u^{11}-\frac{4563}{4} u^{12}+\frac{126411873}{5120} u^{13}\nonumber\\
   &-\frac{221629365}{2048} u^{14}+\frac{648558969807 }{5242880}u^{15}
 \; .
\end{align}
This can be compared with the data of the effective Polyakov loop action and the full theory. 
In the effective theory we compute
\begin{eqnarray}
\label{eq:dels}
 \Delta S(\beta)&=&\frac{1}{T^4}\sum_{n} \left( \left. \frac{T}{V} \frac{d \ln Z}{d \lambda_n}\frac{d\lambda_n}{d u}\right|_{\{\lambda_n=\lambda_n[N_\tau,u]\}}
               -\left.  \frac{T}{V} \frac{d \ln Z}{d \lambda_n}\frac{d\lambda_n}{d u}\right|_{\{\lambda_n=\lambda_n[N_s,u]\}} 
\right)\frac{du}{d\beta}\\ 
             &=&3N_\tau^4  \left( \frac{1}{N_\tau}\left. \erw{R_1(\{\lambda_i\})}\frac{d\lambda_1}{d u}\right|_{\lambda_1=\lambda_1[N_\tau,u]}
               - \frac{1}{N_s}\left. \erw{R_1(\{\lambda_i\})}\frac{d\lambda_1}{d u}\right|_{\lambda_1=\lambda_1[N_s,u]} \right)\frac{du}{d\beta}\nn\\
                          &&+3N_\tau^4\sum_{n=2}  \left( \frac{1}{N_\tau} \left. \erw{R_n(\{\lambda_i\})}\frac{d\lambda_n}{d u}\right|_{\lambda_i=\lambda_i[N_\tau,u]}
               - \frac{1}{N_s}\left. \erw{R_n(\{\lambda_i\})}\frac{d\lambda_n}{d u}\right|_{\lambda_i=\lambda_i[N_s,u]} \right)\frac{du}{d\beta}\; .\nn
\end{eqnarray}
 where all expectation values are calculated on a $N_s^3$ lattice and $\langle R_n\rangle=1/(3N_s^3)d\ln Z/d\lambda_n$.
In the one-coupling theory the last equation reduces to the first line with
\begin{equation}
 R(\lambda_1)=\frac{1}{3N_s^3} \sum_{<i,j>} \frac{ 2 \re (L_i L^\dag_j)}{1+2\lambda_1 \re(L_iL^\dag_j)}\; .
\label{eq:R1L}
\end{equation}
Again an expansion in the limit of small $\lambda_1$ provides a good check of the results. 
The expectation value of $R$ has a simple form in this limit, where it is dominated   
by the nearest-neighbour contribution between adjacent points ($i,j$) on the lattice,
\begin{equation}
 \erw{R_1(\lambda_1)}\approx 2\Big(4\lambda_1^3+44\lambda_1^5 + O(\lambda_1^6)\Big)\; .
\end{equation}
In this approximation finite volume corrections have been neglected. 

We can now appreciate the 
difference to the situation for the string tension.  Comparing eqs.~(\ref{eq:dels}), (\ref{eq:ssc}), we see
that they both have the same trivial $N_\tau^4$-dependence as a prefactor. Any other dependence
on $N_\tau$ in eq.~(\ref{eq:dels}) is contained in the $\lambda_n$. Once again mixed polynomials in the 
$\lambda_n$ are needed to reproduce higher coefficients of the 4d strong coupling expansion, but the 
power counting in $u$ can be based on that of the $\lambda_n$ directly, without cancellations of
$N_\tau$-dependences as in the case of the string tension. The reason is that in the derivation of
the effective theory the "observable" computed as a strong coupling series is the effective action itself, and thus the partition function. These analytic considerations are borne out by numerical simulations.

\begin{figure}[t]
 \includegraphics[width=0.5\textwidth]{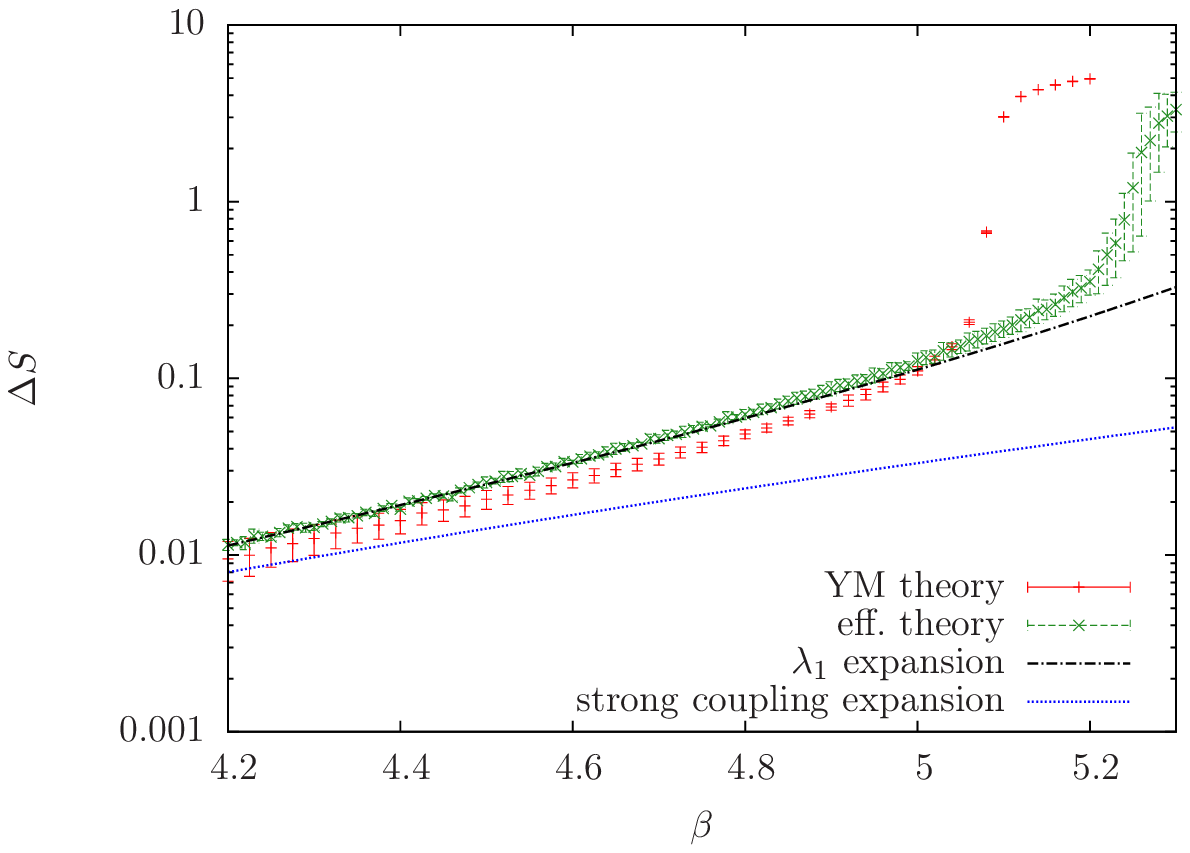}
\includegraphics[width=0.5\textwidth]{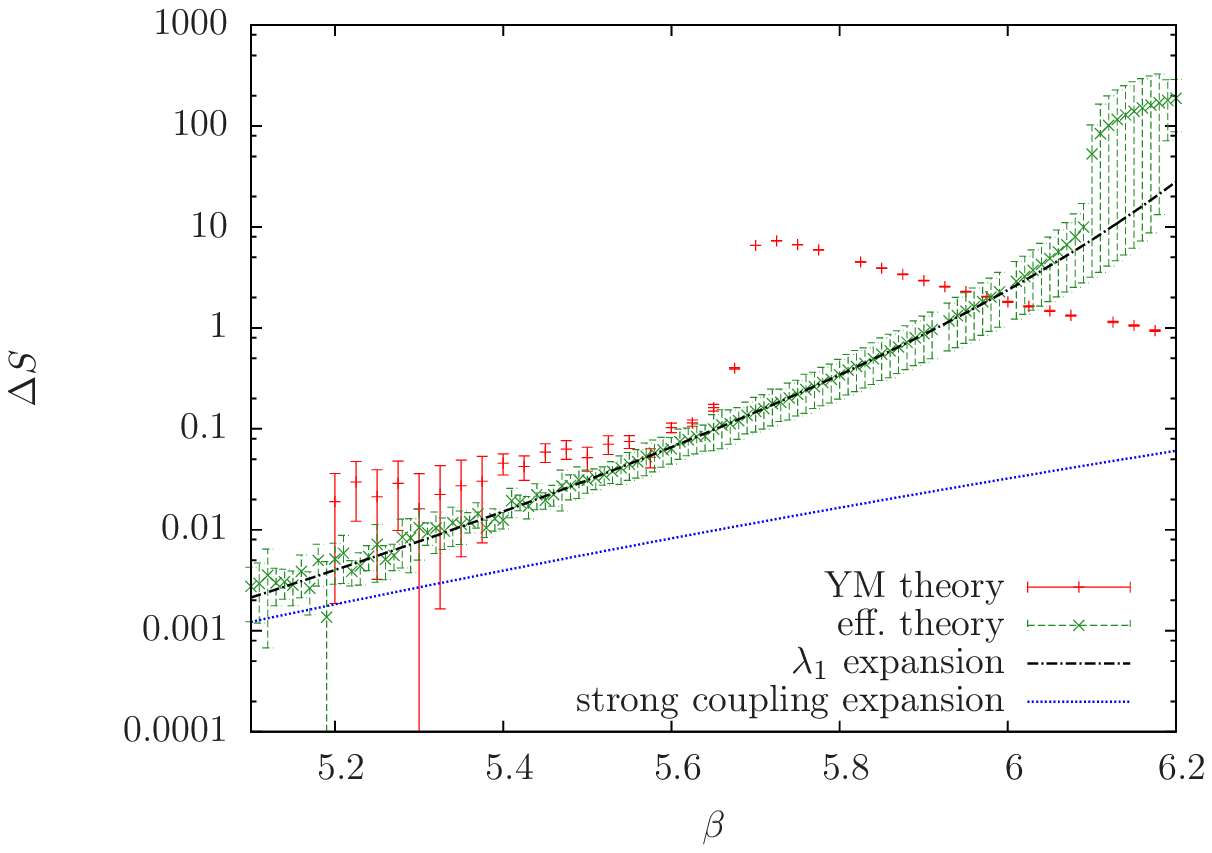}
 \caption[]{$\Delta S$ obtained in simulations of the full Yang-Mills theory 
 and the effective theory. The left panel shows a lattice size of $8^3\times 2$ and $8^3$ respectively;
 the right panel a lattice size of $16^3\times 4$ and $16^3$. A systematic error is included as the difference of the $O(u^9)$ and $O(u^{10})$ truncation in eq.~(\ref{eq:couplings1}). Also shown is 
the 
result of the small $\lambda_1$ expansion and the strong coupling expansion \cite{Langelage:2010yn}.
}
 \label{fig:deltas}
\end{figure}

Unfortunately,
the numerical determination of this quantity  is quite difficult due to large cancellations in the subtraction, which implies a small signal-to-noise ratio. Particularly in the region of strong coupling (small $\beta$), 
$\Delta S$ drops exponentially causing a corresponding growth of the signal to noise ratio. For
an overview and references, see \cite{oprev}.
It is then 
hard to bridge the gap between analytic strong coupling predictions \cite{Langelage:2010yr}
and simulations. In this regime the effective theory turns out to be very useful.
Numerical results for $\Delta S$ as a function of the gauge coupling 
are shown in figure \ref{fig:deltas} for $8^3\times 2$ and $16^3\times 4$ lattices.
The data demonstrate the advantage of the effective theory. Due to the larger statistics obtained in simple three dimensional simulations a 
much higher precision can be achieved than in the full theory. This allows to extend the considered 
$\beta$-range towards smaller values, in particular for higher $N_\tau$ where this is impossible in the full theory. This corresponds to the region of lower temperatures.
We observe excellent quantitative agreement over a wide $\beta$-range. Deviations between full and effective theory predictions only set in at the deconfinement phase transition, with
$\beta^c< \beta_c^\mathrm{eff}$, cf.~table \ref{tab:su3_betas}.

In figure  \ref{fig:deltas} we also compare with the fully analytic results of the strong coupling expansion and the expansion in small $\lambda_1$ within the effective theory.
In a large $\beta$-range the small $\lambda_1$ expansion gives an excellent description.
This indicates that for bulk quantities the short range interactions are dominant in the region well below 
the phase transition. 
Note that the $\lambda_1$-expansion is a much better 
approximation than the strong coupling series for the 
pressure in \cite{Langelage:2010yn}. 
These two results converge, of course, in the strong coupling limit. 
At larger $\beta$ the use of the effective theory entails non-perturbative resummations compared
to the straightforward strong coupling expansion, leading to an improved convergence towards the full theory.

\subsection{Validity of the effective action for thermodynamics and phase transitions}

While the critical couplings for the deconfinement transtion, table \ref{tab:su3_betas}, have already been 
determined in \cite{Langelage:2010yr}, we would like to discuss here why the effective theory works 
so well for this observable. In statistical mechanics, a standard observable to locate a phase boundary is 
the generalised susceptibility of an observable $O(\bx)$, 
\beq
\chi_O=\int d^3x \left(\langle O(\bx) O(0)\rangle -\langle O(\bx) \rangle \langle O(0)\rangle \right).
\label{susc} 
\eeq
At a phase transition fluctuations are maximal, hence
the peaks of susceptibilities define (pseudo-) critical
couplings, whose finite size scaling moreover contains information about the order and universality
class of the transition. 
The important observation is that, despite the integration over all distances, eq.~(\ref{susc}) is a local observable for any theory with a mass gap. The correlators decay exponentially with distance,
\beq
  \langle O(\bx) O(0)\rangle\sim \sum_n c_n^2e^{-E_n|\bx|}\;,
\eeq
with some energy eigenvalues $E_n$ and matrix elements $c_n$,
such that the integral is dominated by the contact and short distance contributions. 
Moreover, at phase transitions
the correlation length of a system either diverges (second order) or is maximal (first order and crossover), 
which implies that any scales smaller than the correlation length play either no or only 
a suppressed role. On the other hand, the behaviour of the correlation length is dictated by the 
symmetries and dimensionality of the
theory. 
A similar reasoning applies for bulk thermodynamic quantities, which are derived from the 
partition function. The non-trivial quantity to compute in this case is the action, which again is local in the
sense that couplings over larger distances are exponentially suppressed. 
Thus, a local effective action with the correct symmetries is capable to provide a good description of bulk thermodynamic quantities as well as phase transitions, even though it might be inaccurate for specific correlation functions or
the spectrum of the theory.

\section{Conclusions}

We have systematically studied the predictive power of a three-dimensional effective Polyakov loop theory for Yang-Mills on the lattice, which has been derived previously by means of a strong coupling 
expansion. The effective theory has an infinite tower of interactions, with coupling between loops at 
all distances, of which only the first few are known analytically.
Here we have tested the simplest version of the effective theory with just one
(resummed) nearest neighbour coupling. Generally the accuracy of effective theory 
predictions depends on the observable where we distinguish to classes: observables characterised by 
explicit length scales, such as correlation functions, and bulk thermodynamic quantities based on the 
partition function or its local derivatives.

The description of correlation functions is found to be quantitatively accurate over short 
lattice distances only, $R/a\approx 0-2$. This is to be expected, since the number of neglected couplings
increases rapidly with distance and the long-range interactions in the effective theory become increasingly important. 
The problem becomes more pronounced as the lattice spacing gets finer. Correlation functions at larger
distances turn out to be systematically underestimated in this particular effective theory, resulting in
an overestimate of the corresponding mass scales. In particular, the temperature dependent effective 
string tension extracted from the free energy of a static quark anti-quark pair is significantly too large
close to the deconfinement transition.

On the other hand, bulk thermodynamic quantities like the equation of state and susceptibilities are 
quantitatively well described when approaching the deconfinement transition. This is because they 
are based on the partition function and thus the effective action itself, which becomes ultra-local in the 
continuum limit. Thus the effective theory is particularly useful for an economic determination of the
phase structure of the underlying full theory. Because of the numerical ease with which accurate results can be 
obtained, the effective theory is superior for a description of the equation of state in the low temperature
regime.
Finally, these conclusions should carry over to the 
effective action describing dynamical QCD, derived by means of a hopping expansion \cite{fromm}, and
its application to finite density phase transitions \cite{fromm2}. 

\section*{Acknowledgements}
G.B.~and O.P.~are supported by the German BMBF, No. 06FY7100. J.L.~is supported
by SNF grant 200020-137920.

\newpage
\begin{appendix}
\section{The couplings of the effective action}
The nearest neighbour interaction is parametrized by the coefficient $\lambda_1$.
In this work we have employed the following series,
\bea
\lambda_1(u,N_{\tau}=2)&=&u^2\exp\bigg[
   2\bigg(4u^4 + 12u^5 - 18u^6 - 36u^7 \nonumber\\
		&&\hspace*{2.0cm}
   + \frac{219}{2}u^8 + \frac{1791}{10}u^9 + 
      \frac{830517}{5120}u^{10}+\ldots\bigg)\bigg],\nonumber\\
 \lambda_1(u,N_{\tau}=4)&=&u^4 \exp \bigg[4 \bigg(4 u^4+12 u^5-14 u^6-36 u^7\nonumber\\
		&&\hspace*{2.0cm}
 +\frac{295 u^8}{2}+\frac{1851 u^9}{10}+\frac{1035317 u^{10}}{5120}+\ldots\bigg)\bigg],\nonumber\\
\lambda_1(u,N_{\tau}\geq6)&=&u^{N_\tau}\exp\bigg[N_{\tau}\bigg(4u^4+12u^5-14u^6-36u^7\nonumber\\
		&&\hspace*{2.0cm}
		+\frac{295}{2}u^8+\frac{1851}{10}u^9+\frac{1055797}{5120}u^{10}+\ldots\bigg)\bigg]\;.
		\label{eq:couplings1}
\eea

The coupling for next-to-nearest neighbours at distance $R/a=\sqrt{2}$ is
\bea
\lambda_2(u,N_{\tau}=2)&=& u^4\bigg(2u^2 + 6u^4 + 31u^6+\ldots\bigg)\; ,\nonumber\\
\lambda_2(u,N_{\tau}=4)&=& u^8\bigg(12u^2 + 26u^4 + 364u^6+\ldots\bigg)\; ,\nonumber\\
\lambda_2(u,N_{\tau}=6)&=& u^{12}\bigg(30u^2 + 66u^4+\ldots\bigg)\; ,
\label{eq:couplings2}
\eea
where the leading coefficient is given by $N_\tau(N_\tau-1)$ for all $N_\tau$.

The next-to-nearest neighbour interactions at distance $R/a=2$ is  
denoted by $\lambda_3$ and has the leading contribution
\begin{equation}
 \lambda_3(u,N_\tau)=4N_\tau u^{2 N_\tau+6}\; .
\end{equation}
\end{appendix}


\end{document}